\documentclass{ws-procs9x6}

\begin{document}

\title{Surface reheating as a new paradigm}

\author{Anupam Mazumdar}

\address{The Abdus Salam, \\
ICTP, Strada Costiera-11, 34100, Trieste, Italy.\\
CHEP, McGill University, 3600-University Road, Montreal, H3A 2T8, Canada}

\author{Kari Enqvist $^{a,b}$, and Shinta Kasuya $^{b}$}

\address{$^{A}$Department of Physical Sciences,\\
P. O. Box 64, FIN-00014, University of Helsinki, Finland.\\
$^{B}$ Helsinki Institute of Physics, P. O. Box 64, FIN-00014,\\
University of Helsinki, Finland.}


\maketitle

\abstracts{In this talk we briefly review the standard idea of reheating
and then present a new paradigm of reheating the Universe through
surface evaporation.}

\section{Introduction}

Finite temperature effects are important to be considered in the
early Universe. Given the fact that there is no direct evidence for
the particle content in the early Universe beyond the Standard Model (SM) 
of electroweak scale $\sim 100$~GeV, and in cosmology 
there is no direct evidence of thermal history beyond the era of Big Bang 
Nucleosynthesis (BBN) $\sim 1$~MeV, it becomes a challenging 
problem understanding the physics of the very early Universe. 
In spite of this there are numerous observational hints which suggests
the extension of the physics beyond the SM. It is also paramount to keep 
in mind that in an expanding Universe the local thermodynamical equilibrium 
can be achieved only if the particle interactions, or at much higher energy 
scales string interactions, or interactions between non-perturbative objects 
such as D-branes are first of all well known, and then their respective 
interaction rates follow: $\Gamma_{i}(T)\geq H(T)$. A simple example will 
illustrate our point. For $2\rightarrow 2$ particle interactions the 
scattering rate is given by $\Gamma\sim \alpha^2T$ ($\alpha$ is a 
coupling constant), which becomes smaller than $H\sim T^2/M_{P}$ (we 
use reduced Planck scale $M_{P}=2.436\times 10^{18}$~GeV) at sufficiently 
high temperatures. It was noticed that elastic $2\rightarrow 2$ processes 
maintain thermal equilibrium typically only up to $T_{max}\sim 10^{14}$~GeV,
while chemical equilibrium is lost already at $T\sim 10^{12}$~GeV
\cite{enqvist90}. In $N=1$ supergravity the situation is completely 
different where the temperature of thermal bath must not exceed 
$10^{10}$~GeV, which we shall discuss below.

In this talk we describe a new paradigm of reheating the Universe,
which is known as surface reheating \cite{enqvist02}. We will
highlight why this mechanism is interesting and may occur in a wide class 
of field theories.


\subsection{Standard lore of reheating and constraints}

Here we quickly review the standard lore of reheating the Universe.
It is commonly believed that inflation is one of the most 
promising early Universe paradigm, which besides
explaining homogeneous, flat and isotropic Universe, also explains 
the seed mechanism for galaxy and large scale structure formation.
The inflaton might be a gauge or non-gauge singlet,
and could also provide a non-vanishing dominating energy 
density which leads to quasi-de-Sitter expansion of the Universe.
The inflaton decays after the end of inflation when it starts 
oscillating about its minimum. The decaying inflaton into a pair
of fermions can reheat the Universe with a relativistic 
thermal bath of temperature \cite{kolb} 
\begin{equation}
T_{rh}=\left(\frac{90}{\pi^2 g_{\ast}}\right)^{1/4}\sqrt{\Gamma_{\phi}
M_{\rm P}}=0.3\left(\frac{200}{g_{\ast}}\right)^{1/4}\sqrt{\Gamma_{\phi}
M_{\rm P}}\,.
\end{equation}
In the above $g_{\ast}$ is the number of relativistic degrees of freedom
at $T_{rh}$ and $\Gamma_{\phi}=\alpha m_{\phi}\sim \alpha H_{inf}$ is the 
inflaton decay rate where $m_{\phi}$ denotes mass of the inflaton, and 
$H_{inf}$ denotes the scale of inflation in many inflationary models. 
The process of thermalization is quite complicated and it may not happen 
instantly \cite{kolb}.

The reheat temperature must be above MeV in order to keep the success
of the BBN. In particularly supersymmetric theories there is also
an upper bound on reheat temperature. The relativistic thermal bath
generates gravitinos thermally \cite{ellis}, and non-thermally
during inflaton oscillations \cite{maroto}. The gravitino interactions 
with matter are Planck mass suppressed, e.g. the helicity $\pm 3/2$ 
mode decays into gauge bosons and gauginos through dimension $5$-operator 
with a life time 
$\tau_{3/2\rightarrow A_{\mu}\lambda}\sim M_{\rm P}^2/m_{3/2}^3$. Gravitino
in a gravity mediated supersymmetry breaking scenario gets a mass of
order $\sim 100$~GeV. This means that they decay after BBN era. The 
over produced gravitinos inject enough entropy to ruin the success of the
BBN \cite{sarkar}. The bound on reheat temperature comes out to be
\begin{equation}
T_{rh}\leq 10^{9}\left(\frac{m_{3/2}}{100~{\rm GeV}}\right)^{-1}
~{\rm GeV}\,.
\end{equation}
It turns out that in high scale inflation models it is hard 
to satisfy this bound on reheat temperature unless the coupling $\alpha$ is
small. In particular the string motivated inflation models where the string 
scale is close to the grand unification scale $\sim 10^{17}$~GeV, and the 
inflaton sector couples to the matter sector via string coupling,  
the problem gets severe, see e.g. \cite{panda}. In general this problem 
is dubbed as {\it gravitino problem}, and only late thermal inflation 
\cite{lythstewart}, or low scale inflation can be the solution, i.e. 
\cite{anu}. Similar problem arises with moduli fields appearing from
string theory. 


\subsection{Surface Reheating}

A novel way to avoid the gravitino and other moduli problems is reheating
via the surface evaporation of an inflatonic soliton. Compared with the
volume driven inflaton decay, the surface evaporation naturally
suppresses the decay rate by a factor:
${\it area}/{\it volume} \propto L^{-1}$
where $L$ is the effective size of an object whose surface is evaporating.
The larger the size, the smaller is the evaporation rate, and therefore
the smaller is the reheat temperature.

Reheating as a surface phenomenon has been considered
\cite{enqvist02} in a class of chaotic inflation models
where the inflaton field is not real but complex. As the inflaton
should have coupling to other fields, the inflaton mass obtains 
radiative corrections resulting in a running inflaton mass with a potential
\begin{equation}
    \label{qpotr}
    V = m^2 |\Phi|^2
    \left[ 1 + K\log\left(\frac{|\Phi|^2}{M^2}\right)\right ]\,,
\end{equation}
where the coefficient $K$ could be negative or positive, and $m$ is
the bare mass of the inflaton. The logarithmic correction to
the mass of the inflaton is something one would expect
because of the possible Yukawa and/or gauge couplings to other fields.
Though it is not pertinent, we note that the potential Eq.~(\ref{qpotr})
can be generated in a supersymmetric theory if the inflaton has a
gauge coupling \cite{enqvistanu} where
$K\sim -(\alpha/8\pi)(m_{1/2}^2/m_{\widetilde{\ell}}^2)$,
where $m_{1/2}$ is the gaugino mass and $m_{\widetilde{\ell}}$ denotes the
slepton mass and $\alpha$ is a gauge coupling constant. It is also
possible to obtain the potential in a non-supersymmetric 
(or in a broken supersymmetry) theory, provided the fermions 
live in a larger representation than the bosons. In this latter 
situation the value of $K$ is determined by the Yukawa coupling 
$h$ with $ K =-C({h^2}/{16\pi^2})$, where $C$ is some number.

As long as $|K| \ll 1$, during inflation the dominant contribution to the
potential comes from $m^2|\Phi|^2$ term, and inflationary slow roll
conditions are satisfied as in the case of the standard chaotic model.
COBE normalization then implies $m\sim 10^{13}$~GeV. If $K < 0$,
the inflaton condensate feels a negative pressure and it is
bound to fragment into lumps of inflatonic matter \cite{enqvist02}. 
Moreover, since the inflation potential Eq.~(\ref{qpotr}) respects 
a global $U(1)$ symmetry and since for a negative $K$ it is shallower 
than $m^2|\Phi|^2$, it admits a $Q$-ball solution, see \cite{enqvistanu}. 
The main idea behind this mechanism is that the fermions coupled to the 
inflaton $h\phi\bar\psi\psi$ decays only through the surface of the 
inflatonic $Q$-ball. The fermion production is blocked by Pauli blocking 
inside the $Q$-ball \cite{cohen}.

The inflatonic $Q$-balls can be created with a size $R\sim |K|^{-1/2}m^{-1}$
when the inflaton oscillates around its minimum. The quantum fluctuations in
the inflaton grow non-linear because of the self-coupling by virtue of 
the Logarithmic correction. The evaporation rate of the fermions can be
given by \cite{enqvist02}
\begin{equation}
\Gamma_Q = \frac{1}{Q}\frac{dQ}{dt}
        \simeq \frac{1}{1.8 |K|^{3/2}}
                \left(\frac{m}{M_{\rm P}}\right)^2 m\,.
\end{equation}
Note that the decay rate is determined by the ratio
$m/M_{\rm P} \simeq 10^{-6}$, which is fixed by the anisotropies
seen in the cosmic microwave background radiation. Even
though we are in a relatively large coupling limit $h\sim 1$, 
the decay rate mimics that of a Planck suppressed interaction of the 
inflatonic $Q$-ball. The reheat temperature turns out to be 
$T_{rh}\sim 10^{8}|K|^{-3/4}$. The value of $K$ depends on the 
nature of the inflaton coupling, but for relatively small value $|K|\sim 0.1$,
we note that we obtain a reheat temperature which can avoid gravitino
and moduli problems.



\end{document}